# Seismic wave shield using cubic arrays of split-ball resonators

Bogdan Ungureanu,[1, a)] Younes Achaoui,[2] Stéphane Brûlé,[3] Stefan Enoch,[3] Richard Craster,[1] and Sébastien Guenneau[3]

[1)]*Department of Mathematics, Imperial College London, London, UK*
[2)]*FEMTO-ST, University of Franche-Comté, Besancon, France*
[3)]*Aix Marseille Univ, CNRS, Centrale Marseille, Institut Fresnel, 13013 Marseille, France*

(Dated: 24 March 2019)

Metre size inertial resonators located in the ground have been theoretically shown to interact with a seismic wave (attenuation, band gaps) to enable protection of surface structures such as buildings[1]. The challenge for Civil Engineering is to both reduce the size of these resonators and to increase their efficiency. Here we explore steel spheres, connected to a concrete bulk medium, either by a coating of rubber, or rubber and steel ligaments, or air and steel ligaments. We show that for a cubic lattice periodicity of 1 metre, we achieve stop bands in the frequency range 14 to 20 Hz; by splitting spheres in 2 and 8 pieces, we tune down the stop bands frequencies and further increase their bandwidth. We thus demonstrate we are able to provide a variety of inertial resonators with stop bands below 10 Hz i.e., in the frequency range of interest for earthquake engineering.

Several research groupings have proposed different routes towards seismic metamaterials with deeply subwavelength local resonances overcoming the usual limitation of narrow frequency band gaps. For surface waves, propagating in sedimentary soils, it is enough to consider a periodic assembly of stiff pillar inclusions (a nomenclature of the different categories of metamaterials with seismic applications is proposed in[2]) clamped to a bedrock to achieve a shielding effect from 0 to 20 Hz[3]. However, pillars usually do not exceed 50m in depth, mainly for economic reasons with current civil engineering technology, so other solutions need be explored when the geotechnical and/or seismological bedrock is far beneath the surface[4]. In this Letter, we will introduce inertial resonators (falling in the category of Seismic Metamaterial and more precisely in the subcategory of Buried Mass Resonator (BMRs)), which can stop elastic waves over larger frequency bandwidths than in our previous work on steel spheres connected to a concrete bulk via thin steel ligaments[1]. These new inertial resonators take the form of steel split balls which are otherwise connected to a concrete bulk and could be implemented as shown in fig. 1. The advantage of using steel split balls rather than simply steel balls is that we achieve even lower frequency stop bands, which are moreover wider. In addition to other types of devices such as seismic insulators, tuned mass dampers, etc. the proposed split-ball resonators (SBRs), which are a type of Buried Mass Resonators (BMRs), may concern sensitive structures such as buildings and dams but also any type of structure afterwards, in addition to existing devices such as dampers, seismic isolators, etc.; for instance, SBRs could be installed in- side soft soil, rocky ground or concrete basement.

Analyzing the band structures associated with the propagation wave field trough periodic media is a new approach in civil engineering; this draws upon the wealth of knowledge in photonics and optics regarding wave control in those fields. This approach unveils the intimate relation between the wave vector and the wave frequency which relates the spatial and temporal characteristics of the wave motion in the ground. The adoption of band structures and the underlying wave control concepts is one of the main advances of this last decade in civil engineering because it relaunches studies on soil-structure interaction, on more precisely, on all kind of structured soil. This dispersion relation is very rich in terms of information and gives us the possibility to identify band gaps, more precisely frequency intervals whereby propagating waves are forbidden to propagate within an infinite periodic medium and so seismic waves are attenuated when propagating though a finite piece of a such structured medium.

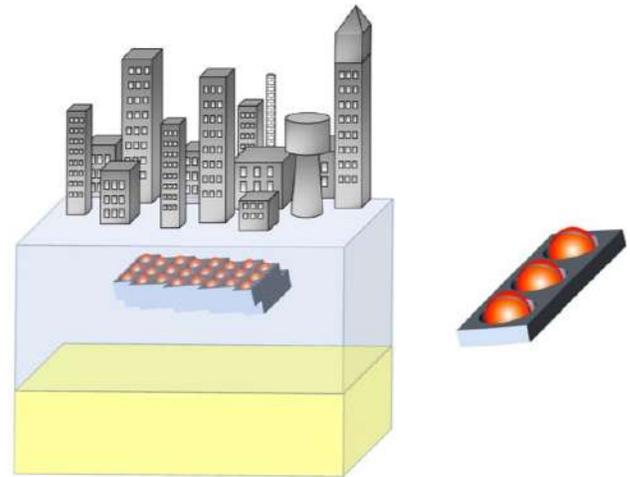

FIG. 1. Schematic, not to scale, representation of a seismic wave shield consisting of a periodic array of split-ball resonators placed underneath the foundations of a large civil infrastructure. Such inertial resonators shield wavelengths much larger than their typical size thanks to low frequency stop bands associated with local resonances.

[a)]Electronic mail: b.ungureanu@imperial.ac.uk

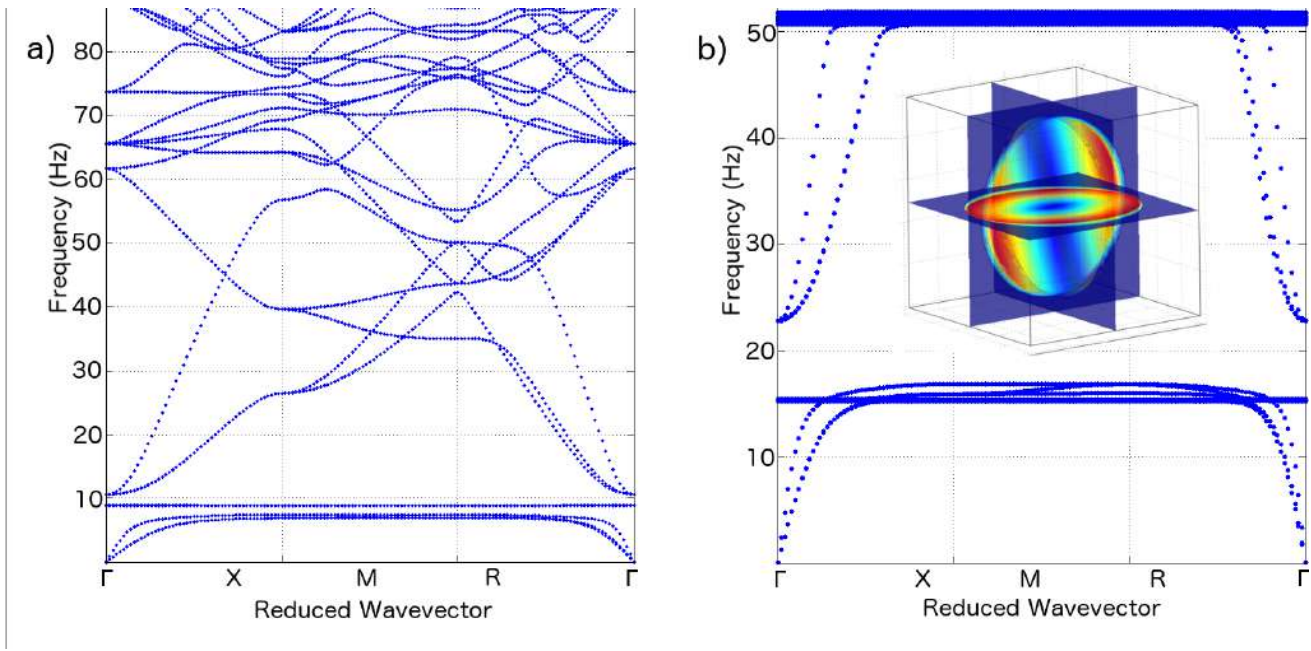

FIG. 2. Band diagrams for an elastic wave propagating within a periodic structure (1 metre in sidelength) made of concrete (host medium - with density $\rho = 2.4\times10^3$ kg m$^{-3}$ and Youngs modulus E = 17x10$^6$ Pa) and inertial resonators (of diameter 0.74m) connected to the propagation medium via 6 steel (density $\rho = 7.870\times10^3$ kg m$^{-3}$ and Youngs modulus E = 200x10$^9$ Pa) ligaments 0.02 m in diameter and 0.03 m in length (left figure) and same with ligaments replaced by a coating with rubber property (a polymeric material with density $\rho = 1.2\times10^3$ kg m$^{-3}$ and Youngs modulus E = 1.1x10$^9$ Pa) (right figure). Vertical axis is the Bloch wave frequency (Hz) and horizontal axis the Bloch wavenumber (m$^{-1}$).

The quality of the band gaps for the seismic protection is given by the width of the frequency range (which in the seismic case should be very low frequencies, typically a frequency interval below 10 Hertz with considerable width). However, it is also important that the depth of the imaginary part of the wave number spectrum which describes the level of attenuation be as large as possible.

We have shown in previous work that band gaps can be obtained with the help of Bragg scattering, local resonance, a dynamic negative Poisson coefficient[5] (which has been coined auxetic seismic metamaterials) etc. Bragg scattering occurs due to the periodicity of a material or structure[6], where waves scattered at the interfaces cause coherent destructive interference, effectively canceling the propagating wave. Using local resonance, which leads to very dispersive properties of the metamaterials means transferring the vibrational energy to a resonator[1]. These eigenmodes of the resonant structure are very precisely described by the band diagrams. For elastic band gaps the concept of local resonance should be considered within the framework of periodic structures to obtain metamaterials properties enabling molding wave propagation through large scale bulks. Low frequency band gaps due to the local resonances are used in order to attenuate waves propagation through continuous structures[7–9].

The first in-situ experiment of a deflection shield for seismic waves was led by the Menard company in collaboration with the Fresnel Institute near the Alpine city of Grenoble in 2012. A Rayleigh wave generated by a source at frequency of 50 Hz within the stop band of a large scale phononic crystal unveiled some similarities between acoustic waves in plates and surface seismic waves and this prompted researchers to envision large-scale analogs of acoustic metamaterials, called seismic metamaterials (and in particular SSM). Experimental results have validated the possibility to prohibit frequencies around 50 Hz i.e. at the upper frequency range of interest as suggested by the idealized single degree of freedom oscillator (SDOFO with oscillators) model, carried out for free vibrations, and further paved the way to explore locally resonant subwavelength structures that might improve such Bragg-based seismic shields. Another experimental step towards seismic metamaterials with subwavelength stop bands has been achieved with forests of trees acting as the resonators[10], (this type of seismic metamaterials is called Above-Surface Resonator, ASR).

The present Letter presents our latest results on new classes of seismic metamaterials, (Buried Mass Resonators), based on subwavelength inclusions that can change the characteristics of surface Rayleigh waves. These can be achieved by engineering the frequency stop bands dedicated to seismic waves by converting the latter to evanescent waves in order to protect a sensitive area. We numerically demonstrate that seismic waves can be molded by changing the field propagation properties, by placing a cubic array of resonators in the wave propagation path.

Numerical simulations, performed in Comsol Multiphysics, analyze the propagation of bulk waves through a structured medium. Floquet-Bloch boundary conditions are applied on the sides of a periodic cell of side length $a$ = 1 m in order to compute band diagrams:

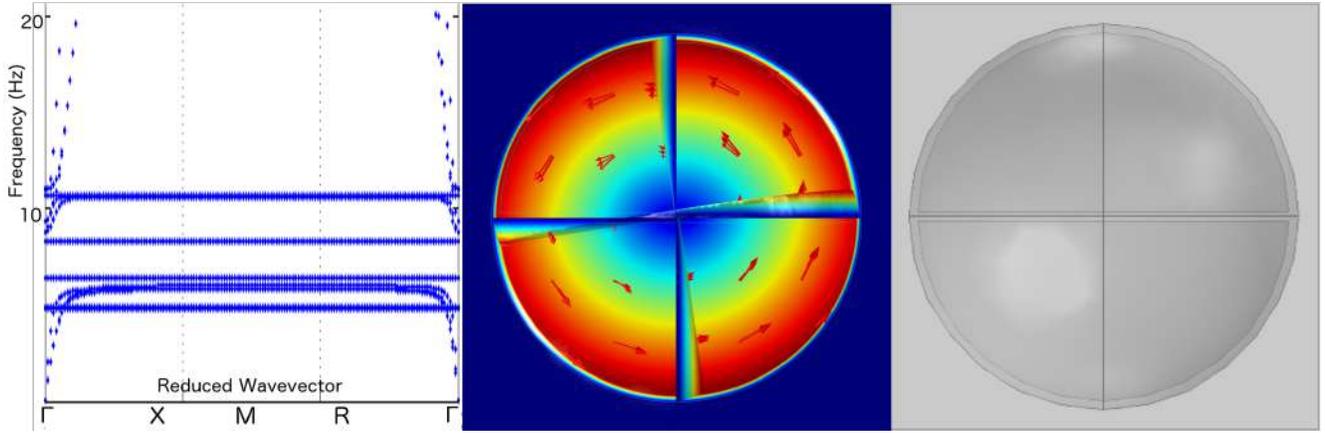

FIG. 3. Band diagrams for elastic wave propagating within a periodic structure with a basic cell 1 metre in sidelength consisting of an steel sphere of diameter 0.74m, split in 2 semispheres connected together by a thin layer of rubber $l_i$ =0.02 m and to concrete (host medium - with density $\rho$ = 2.4x10$^3$ kg m$^{-3}$ and Youngs modulus E = 17x10$^6$ Pa) by a coating of rubber (a polymeric material (Vero Blue) with density $\rho$ = 1.2x10$^3$ kg m$^{-3}$ and Youngs modulus E = 1.1x10$^9$ Pa). Vertical axis is the Bloch wave frequency (Hz) and horizontal axis the Bloch wavenumber in the irreducible Brillouin zone (m$^{-1}$).

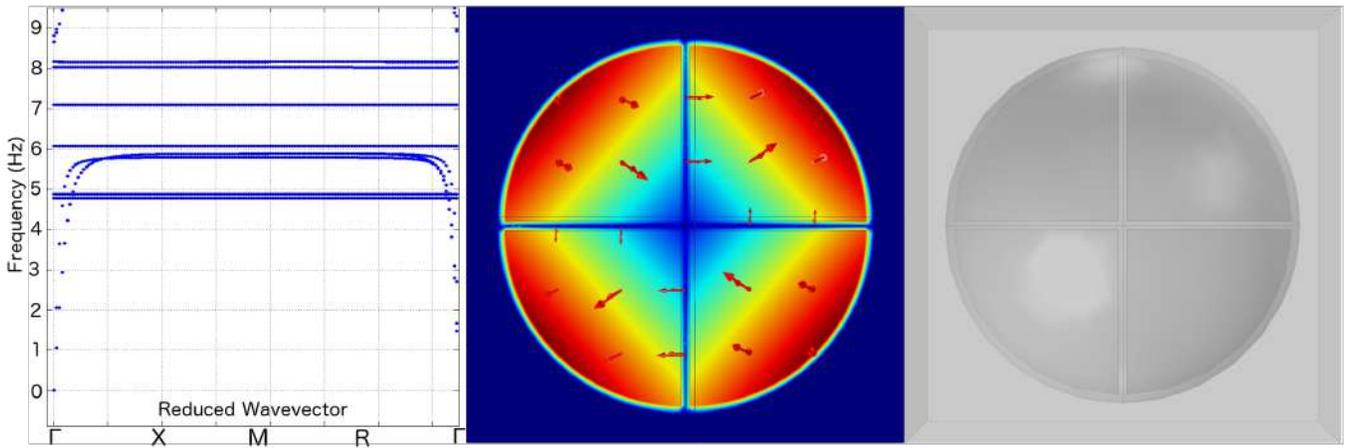

FIG. 4. Same as Fig. 3 for 8 pieces connected together and to concrete (host medium) by a coating of rubber.

For each value of the Bloch wavenumber chosen on the edges the irreducible Brillouin zone (IBZ), which in the present case is a tetrahedron with vertices $\Gamma$ = (0, 0, 0), $X$ = ($\pi/a$, 0, 0), $M$ = ($\pi/a$, $\pi/a$, 0) and $R$ = ($\pi/a$, $\pi/a$, $\pi/a$), an eigenvalue problem is solved that leads to a countable set of eigenvalues associated with Floquet-Bloch waves undergoing a phase shift when passing through the cell; and when the Bloch wavenumber describes the IBZ, the union of spectra forms bands, sometimes revealing so-called stop bands i.e. absence of eigensolutions: these are frequency ranges where waves are disallowed to propagate within the periodic medium. Our objective is to design such stop bands at low frequencies, that is when the wave wavelength is much larger than the size of the periodic cell: in this way we can hope to deflect seismic waves thanks to a subwavelength structure of the soil.

We start from our earlier work in[1], and thus we consider a periodic cell consisting of a block of concrete 1 ×1× 1 m$^3$ in volume, with a 0.8 meter-diameter inner hole hosting a 0.74 meter-diameter steel ball connected to the concrete bulk medium by 6 steel ligaments a coating of rubber. The corresponding band diagrams is shown in the left panel of Fig. 2. We then replace the ligaments by a coating of rubber and report the corresponding band diagram in the right panel Fig. 2. We note a larger stop band in the right panel, which is nonetheless at higher frequencies. We thus need an additional idea to achieve both low frequency and large stop bands. Motivated by the concept of structural interface in dynamic elasticity introduced in[11] we proceed to break up the spheres into smaller resonators connected by soft structured interfaces.

On the band diagram of Fig. 3, where we consider the same rubber coating, we have split the ball in two halves, connected by rubber. We notice the appearance of two flat bands around 5 Hz, associated with extremely narrow stop bands, which should be compared with the large stop band around 20 Hz in Fig. 2, right panel. The origin of these very tiny stop bands is due to the presence of local resonances at those frequencies. These local resonances do not couple strongly enough to open band gaps useful in practice, but the next resonances placed around 7 Hz are responsible for a large band gap that can be seen in the figure 3.

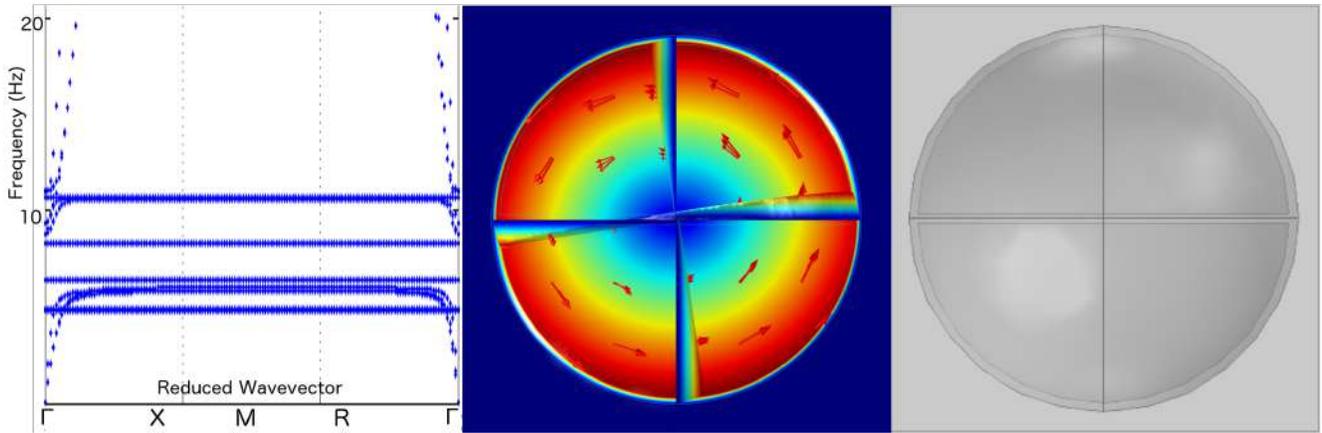

FIG. 5. Same as Fig. 4 for 8 pieces connected together and to concrete (host medium) by 20 evenly spaced steel ligaments of length $l_i$ =0.03 m and diameter $h_i$ =0.02 m.

Since cutting the sphere in two halves has been beneficial, we then proceed with cutting the sphere in 8 parts. The result is shown in Fig. 4, where one can clearly see the stop band opened even further, yet remained within the same frequency range. On the other hand, a strong coupling between local resonances and the continuum may be observed at about 7 Hz where a wide full stop band occurs. Because the compressional swinging and rotational modes are close to each other and the former can easily couple to longitudinal waves and the latter to shear waves. The gap bandwidth is obviously related to the quality factor of the resonators.

Out of curiosity, we finally considered the ball split in 8 parts, when we replace rubber by 20 evenly spaced steel ligaments, see Fig. 5. This leads as one would expect to stop bands at lower frequencies than in Fig. 4 (centered around 2 Hz), and besides the stop band remains fairly large. We would like to stress that we were never able to achieve such a large stop band at such low frequencies before, bearing in mind the periodic cell size is the same as in our former work[1].

We pointed out that by varying the sizes of ligaments, we bring all flat bands in Fig. 5 inside the stop band, thus improving the damping properties of the shield. However, we do not proceed with such numerics in this Letter, as by doing so we would need to adapt the irreducible Brillouin zone (to take into account symmetry breaking). As we proposed in previous work this phenomenon could be obtained also by changing the size of the ligaments and transforming the pure modes of the resonator in to hybrid modes, changing the eigenfrequency values. For a better exploitation of these resonances, we have suggested therein to differentiate the size of the ligaments and this action permits effectively to separate the resonances[3]. This makes it possible to broaden the range of protection and the quality factor gets better, and all this due to the fact that the pure modes, observed in the case of identical ligaments, become semi hybrid modes in the case of ligaments with variable dimensions.

To conclude, by splitting spheres in 2 and 8 pieces, we tuned down the stop bands and further increased their bandwidth. Potential applications of our study are in seismic shields against surface elastodynamic waves with short wavelengths that occur in sedimentary basin.

We hope that our study will foster experimental efforts in meter and decameter scale metamaterials for surface mechanical wave control.

**ACKNOWLEDGMENTS**

Bogdan Ungureanu acknowledges funding of European Union (MARIE SKODOWSKA-CURIE ACTIONS project Acronym/Full Title: METAQUAKENG - Metamaterials in Earthquake Engineering - MSCA IF - H2020).